\documentclass{llncs}
\usepackage{graphicx}
\usepackage{subfigure}
\usepackage{verbatim}

\begin{document}


\mainmatter

\title{ChaMAILeon\\Simplified Email Sharing Like Never Before}

\titlerunning{Lecture Notes in Computer Science}

\author{Prateek Dewan \and Mayank Gupta\and Ponnurangam Kumaraguru}
\institute{Indraprastha Institute of Information Technology, Delhi\\
\email{prateek1014@iiitd.ac.in, mayank.gupta@dce.edu, pk@iiitd.ac.in\\
~\\precog.iiitd.edu.in}}
\maketitle

\begin{abstract}

While passwords, by definition, are meant to be secret, recent trends in the Internet usage have witnessed an increasing number of people sharing their email passwords for both personal and professional purposes. As sharing passwords increases the chances of your passwords being compromised, leading websites like Google strongly advise their users not to share their passwords with anyone. To cater to this conflict of usability versus security and privacy, we introduce ChaMAILeon, an experimental service, which allows users to share their email passwords while maintaining their privacy and not compromising their security. In this report, we discuss the technical details of the implementation of ChaMAILeon.

\end{abstract}

\section{Introduction and motivation}
In the recent past, the practice of \emph{sharing} passwords among teenagers has caught special attention in leading news media all over the world. News agencies like New York Times, China Daily, Zee News India, Times of India and many more have reported how sharing passwords has become a new way of building trust and intimacy amongst teenagers and couples. The Pew Internet and American Life Project found that 30 percent of teenagers who were regularly online had shared a password. The survey of 770 teenagers ages 12 to 17, found that girls were almost twice as likely as boys to share. And in more than two dozen interviews, parents, students and counselors said that the practice had become widespread~\cite{Amanda-Lenhart:2011}.

While some individuals are comfortable with sharing their passwords with their best friends and partners, other individuals who do not wish to share their passwords often land up on a rough side~\cite{Boyd:2012}. People even suffer a break up only because they did not share their passwords~\cite{Kamra:2012}.

The government of India was recently thinking of a code of conduct by which Government employees would use Facebook. The interesting aspect of this ``code'' was that ``the password of the account would be known to others in the department''~\cite{Times:2011}. While ``Don't share your password with anyone!''~\cite{Shen:2008} is one of the most basic rules of staying secure over the Internet, the growing urge amongst teenagers for sharing them is leading to growing security problems and major privacy issues.

To tackle these issues, we came up with the idea of sharing emails in a way the user wants. ChaMAILeon gives users the freedom to share what they want to share and preserve what they don't wish to share from their emails. This service suits the requirements of teenagers who wish to share their passwords with their partners as a sign of trust, but want to maintain their privacy at the same time.

\section{Problem statement}

Sharing passwords is becoming a necessity for a large number of the Internet users today, but what the users don't realize is the implications of them sharing their passwords. Having to share their email password with someone not only raises concerns for an individual's privacy, but also makes them vulnerable to serious risks. Someone knowing your password could change it, delete your account before you even come to know, or even worse, send unwanted emails to your boss, partner or anyone you can think of! The problem arises when you are the one who is blamed for all this, beacuse of the fact that you are the owner of the account. Moreover, an individual asked to share his/her password with their partner or soemone else is not always comfortable in doing so as he/she might have privay concerns and not want to share everything~\cite{Kamra:2012}. On the other hand, not sharing their passwords leads them into quarrels, fights or even break ups; as one of the articles mentioned above suggests. Furthermore, if you are connected to an untrusted network like a cyber cafe or an open network, and wish to access your email, doing so might not be a good idea. Untrusted networks often capture all network traffic and may try to extract your password too. So, building a system to enable users to share their password and not be worried about privacy was our main motivation.

\section{The solution}

\begin{figure}
\fbox{\includegraphics[scale=0.3]{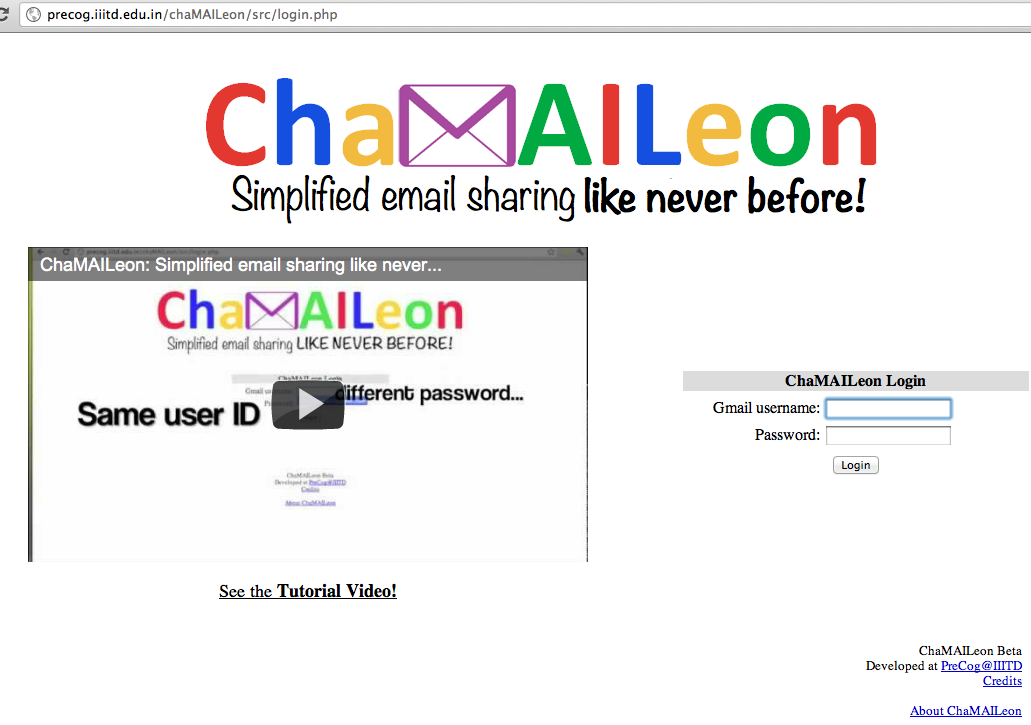}}

\caption{ChaMAILon login page}

\label{ch_homepage}

\end{figure}

\begin{figure}
\fbox{\includegraphics[scale=0.33]{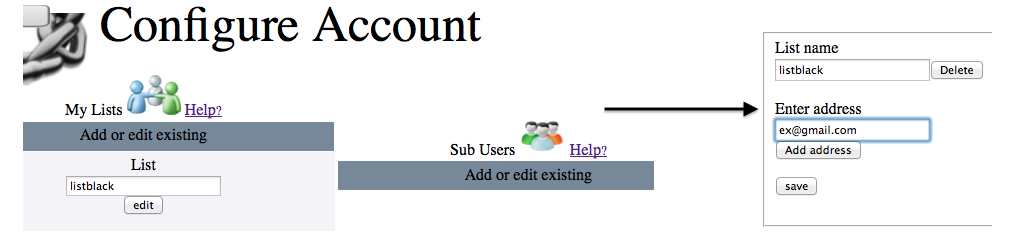}}

\caption{Creating a list}

\label{list}

\end{figure}

\begin{figure}
\begin{center}
\fbox{\includegraphics[scale=0.4]{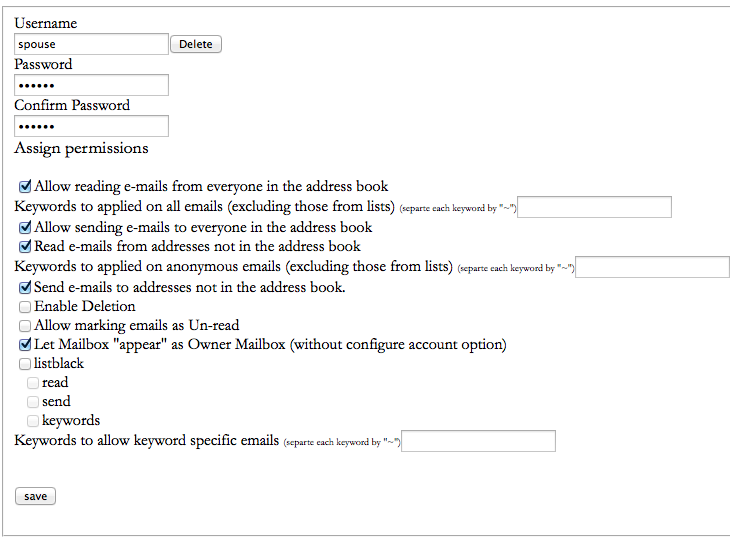}}

\caption{Setting permissions}

\label{subuser}
\end{center}
\end{figure}

To address this problem of sharing passwords while maintaining privacy and security, we came up with an experimental service which allows users to share their emails the way they want. ChaMAILeon allows users to create multiple passwords for their account, and associate a different level of access with each such password.
For example, consider a user with a username \emph{person@gmail.com}
and password \emph{`appleball'}, who wants to share her password with her spouse. However, this user has certain emails in her inbox from her ex-boyfriend (ex@gmail.com), which she does not want her spouse to see, and does not want to delete them either. If she does not share her password with her spouse, it creates doubts between them, but if she does share her password, and her spouse sees those emails, things might not turn out very well.
Ideally, the user would want to share her password in such a way, that when her spouse logs in to her email account, he cannot view emails from \emph{ex@gmail.com}.
This can now actually be done using ChaMAILeon. person@gmail.com logs in to ChaMAILeon (Figure \ref{ch_homepage}) using her actual password `appleball' and goes to the ``Configure Account'' page (Figure \ref{list}). Here, she creates a new list (say \emph{listblack}) and adds the email address \emph{ex@gmail.com} to it.
Now, she creates a sub user (say \emph{spouse}) and creates another password for her email ID, say \emph{`catsanddogs'}. She sets rules allowing read permissions on emails coming from all email addresses except for the ones present in the \emph{listblack} (Figure \ref{subuser}).

\begin{figure}
\centering
\subfigure[User logs in with password `appleball' and sees all emails]{
\includegraphics[scale=0.4]{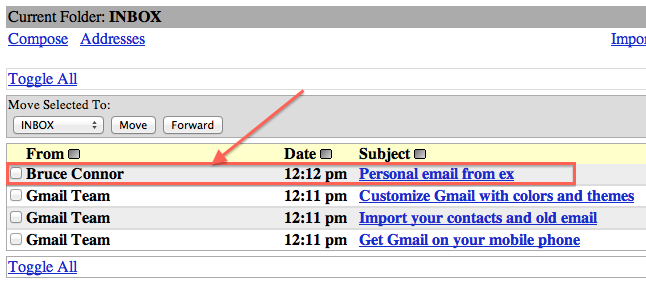}
\label{fig:subfig1}
}
\subfigure[Spouse logs in with password `catsanddogs' and cannot see email sent from `ex@gmail.com']{
\includegraphics[scale=0.4]{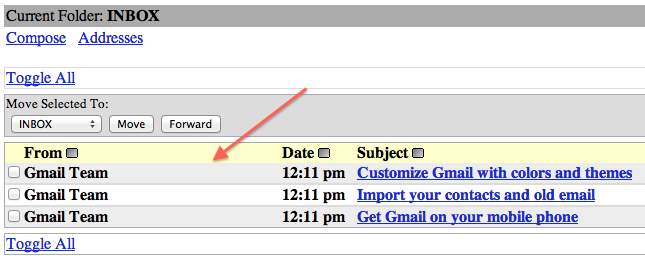}
\label{fig:subfig2}
}

\label{inbox}
\caption[]{Share your password and still maintain your privacy}
\end{figure}

Now, the user tells her spouse that her password is \emph{catsanddogs}. The spouse goes to ChaMAILeon (Figure \ref{ch_homepage}), logs in to the user's account using her email ID 'person@gmail.com' and password `catsanddogs' and gets to view all her emails, except the ones coming from 'ex@gmail.com' (Figure \ref{inbox}). The spouse doesn't even get a hint that some emails (from ex@gmail.com) are missing!

ChaMAILeon provides blocking or allowing emails from specific people and also provides allowing emails containing only certain keywords. All these specifications and preferences can be set by the users to suit their needs.

\section{Setting the preferences}

Currently, ChaMAILeon supports only Google accounts. One can access their Gmail account through ChaMAILeon instead of going to the Gmail website. By logging into ChaMAILeon, users get the ability to access their account with more than one password. Each new password they create can be configured to allow restricted access at different levels to their account. To set up, follow these steps:
\begin{enumerate}
\item Log into your Google Mail account through ChaMAILeon.
\item Go to the ``Configure Account'' option on the top right corner of the Inbox page.
\item Create lists and add email IDs to these lists; you can use them as black lists or white lists. (Optional)
\item Create another password for your account by creating a sub user. (You should share this password and not your actual password)
\item Assign your desired permissions to this sub user and click Save.
\end{enumerate}

\subsection{Technical specifications}

ChaMAILeon has been developed using SquirrelMail, an open source standards-based webmail package written in PHP. We have added on to the standard implementation of SquirrelMail to accommodate sharing features. We also use a MySQL database at the back end to store contacts, lists, passwords (in encrypted form) and other sub user details.

\section{Conclusion}

In this report, we introduced ChaMAILeon, a system which allows users to share their email passwords while maintaining their privacy. The system provides restricted email sharing, by allowing users to create multiple passwords for their email account, where each password allows a different level of access to the user's email. Users can set access preferences and filters based on keywords and senders. The system also supports black listing and white listing for emails coming from a fixed list of email addresses.

\section{Acknowledgements}

We would like to take this opportunity to thank Tarun Bansal, Vartika Srivastava and Monika Singh for giving a try to this idea(as part of their foundations in computer security course) and making us believe that such a system could really work. We would also like to thank Sheethal Shreedhar, who really worked hard and helped us lay the building blocks for this service.

\bibliographystyle{plain}
\bibliography{ch_bib}
\end{document}